\documentclass[12pt]{article}\begin{document}

   \newtheorem{lem}{ \sc Lemma}

   \newtheorem{df}{ Definition }

\newcommand{\eqdef}{\stackrel{\rm def}{=}}

 \newtheorem{te2}{ \sc Theorem}
 \newtheorem{le2}{ \sc Lemma}
 \newtheorem{df2}{ Definition}

\title{\Large  Quadratically integrable geodesic flows on the torus and on the Klein bottle
}
\author{\it Vladimir Matveev\footnote{vmatveev@physik.uni-bremen.de} } \date{}
\maketitle

\begin{abstract}

1. If the geodesic flow of a metric $G$ on the torus $T^2$
is quadratically integrable then the torus $T^2$ isometrically covers
a torus with a Liouville metric on it.

2. The set of quadratically integrable geodesic flows
on the Klein bottle is described.
\end{abstract}

  {\bf \S1.  Introduction}
 Let  $M^2$  be a smooth close surface with
 a Riemannian metric  $G$ on it. The metric allows to canonically identify the
 tangent and the co-tangent bundles of the surface $M^2$. Therefore we have
 a  scalar product and a  norm in every co-tangential plane.

 \begin{df2}  Hamiltonian system on the co-tangent plane with the
 Hamiltonian   $H\eqdef|p|^2$ is called  {\bf  the
 geodesic flow of the metric $G$}.
\end{df2}

 It is known that the trajectories of the geodesic flow project (under the natural
 projection $\pi$, \ $\pi(x,p)\eqdef x$)   in the geodesics.

 \begin{df2}
      A geodesic flow is called {\bf integrable } if it is integrable as the
      Hamiltonian system. That is there exists a function
      $F:T^*M^2\rightarrow R$ such that: {\small
\begin{itemize}
      \item  
      $F$ is constant on the trajectories, \item
      $F$ is functionally independent with $H$.\end{itemize}}

      The function $F$ is called {\bf an integral}.
 \end{df2}

 Recall that two functions ($F$ and $H$ in our case) are  {\it
 functionally independent} if the differentials $dF$ and $dH$ are
 linear independent almost everywhere.

 \begin{df2}
      A geodesic flow is called {\bf linear  integrable }
       is there exists an integral
      $F:T^*M^2\rightarrow R$ such that in a neighborhood of any point the integral
      $F$ is
      given by the formula
       $F(x,y,p_x,p_y)=a(x,y)p_x+b(x,y)p_y$, where $x,y$ are  coordinates
        on the surface,  $p_x, p_y$ are the correspondent momenta, and
        $a,b$ are smooth functions of two variables.
        \end{df2}

 \begin{df2}
      A geodesic flow is called {\bf quadratically integrable } if it is
      not linear integrable and  if
       there exists an integral
      $F:T^*M^2\rightarrow R$ such that in a neighborhood of any point the integral $F$ is
      given by the formula
       $F(x,y,p_x,p_y)=a(x,y)p^2_x+b(x,y)p_xp_y+c(x,y)p_y^2$, where $x,y$ are  coordinates
        on the surface, $p_x, p_y$ are the correspondent momenta, and
        $a,b,c$ are smooth functions of two variables.
        \end{df2}

The set of linear and  quadratically integrable geodesic flows on the closed
oriented surfaces was
completely described in  \cite{BN},
\cite{Koz},  \cite{Kol1}, and \cite{Mat}.
In particular, in \cite{Koz} it was proved that there are no such geodesic
flows  on the surfaces of genus $g>1$ (see also \cite{Ta}).
Linear and quadratically integrable geodesic flows on the
sphere were completely described in \cite{Kol1}.
Quadratically integrable geodesic flow on the torus
were described in \cite{BN}.
Linear integrable geodesic flows on the torus were
described in \cite{Mat}.

The aim of the present paper is to
give an other description of the set of the quadratic
integrable geodesic flows    on the torus and to describe the set
of quadratically  integrable geodesic flows on the Klein bottle.

First recall briefly (following \cite{BN}) the description of quadratically integrable geodesic
flows on the torus.

Suppose $L$ is a positive number. Let $S_1$, $S_L$ be circles, supplied with
 smooth parameters
  $x\in R(\ \mbox{mod} \ 1)$ and
  и $y\in R(\
 \mbox{mod} \ L)$ respectively.

   \begin{df2}   The metric
$(f(x)+h(y))(dx^2+dy^2)$   on the torus  $S_1\times S_L$, where
 $f:S_1\to R$ and  $h:S_L\to R$ are positive smooth non-constant functions,
 is called  {\bf a Liouville} metric.
                                  \end{df2}

\begin{df2} A metric $G$ on the torus  $T^2$ is called {\bf pseudo-liouville}
if there exists a Liouville metric  $G_{Liuv}$ on the torus
$S_1\times S_L$ (for an appropriate  $L$) and  a covering
 $\rho:S_1\times S_L\to T^2$ such that  $G_{Liuv}=\rho^*(G)$. \end{df2}

 \begin{te2}[\cite{BN}]
   The geodesic flow of a metric  is quadratically integrable iff the
   metric
   is pseudo-liouville. \end{te2}

   The theorem describes the set of quadratically integrable geodesic flows
   on the torus.

   We would like to suggest one more description.

\begin{te2}  A metric  $G$ on the torus  $T^2$ is pseudo-liouville
iff there exists a Liouville metric
 $G_{liuv}$ on the torus $S_1\times S_l$ (for an appropriate  $l$)
 and a covering
$\chi:T^2\to S_1\times S_L$ such that  $G=\chi^*(G_{liuv})$.
\end{te2}

Theorem 2 will be proved in \S2.

In other words, for every metric $G$ (on the torus  $T^2$) with the
 quadratically integrable
 geodesic flow there exists a composition covering
 $\rho:S_1\times S_L\to T^2$,\ $\chi:T^2\to S_1\times S_l$
that takes first a Liouville metric to the metric $G$ and then
the metric $G$ to a Liouville metric.

How to specify a metric with the quadratically integrable geodesic flow?
A Liouville metric is specified  by the triple $(L,f,h)$.
It is known that   the finite  coverings of the torus are specified by
$2\times 2$ integer non-degenerate matrixes. Two matrixes ($A$ and $B$)
give equivalent
coverings if     $A=XB$ for a integer matrix $X$, det$(X)=\pm1$.
It is easy to prove that each integer non-degenerate $2\times 2$
matrix is equivalent
(i.e., gives the  equivalent covering) to        the appropriate matrix
$\left(\matrix{1&k\cr0&m\cr}\right)$, where $0<k\le m$.
Two matrixes
$\left(\matrix{1&k_1\cr0&m_1\cr}\right)$ and
$\left(\matrix{1&k_2\cr0&m_2\cr}\right)$ ($0<k_1<m_1,\ 0<k_2<m_2$)
are equivalent if they are equal.

Thus a metric with the quadratically integrable geodesic flow
is specified by the quadruple $(L,f,h,\frac{k}{m})$.

                                    \vspace{.2cm}

In \S3 we describe quadratically integrable geodesical flows on the Klein bottle.

Let  $L$ be positive number, $f,h:R\to R$ be smooth functions such that
function $f$ is non-constant and for any $x,y$ \
 $f(x+\frac{1}{2})=f(x)$, $h(y+L)=h(y)=h(-y)$.

 For every triple $(L,f,h)$  we shall construct  the metric $G_{(L,f,h)}$ on the
 Klein bottle $K_{L}$. The geodesic flow  of the metric $G_{(L,f,h)}$ is quadratically
 integrable.
 If the geodesic flow of a metric on the Klein bottle $K$
 is
 quadratically integrable then there exists a diffeomorphism  $K\to K_L$ that
 preserves metric.

\vspace{.2cm}

   {\bf \S2. Proof of Theorem 2.}

 \begin{le2}     Let  $f:R\rightarrow R$, $h:R\rightarrow R$ be {bounded }
 smooth functions.
 If  for vector
$\bar u=(u_1,u_2)$  and  for any $x,y$
$f(x+u_1)+h(y+u_2)=f(x)+h(y)$, then
$f(x+u_1)=f(x)$ and $h(y+u_2)=h(y)$.  \end{le2}

\vspace{.2cm}

Proof. We have
$f(x+u_1)-f(x)
=h(y)-h(y+u_2)$. Therefore for the appropriate  constant
      $C$ we have 
$f(x+u_1)-f(x)=
C=h(y)-h(y+u_2)$.
  Let us show that $C=0$. Actually,
$f(x+nu_1)=f(x)+nC$. If $C\ne 0$, then   function $f$ is not bounded function.
Proved.

                 \vspace{.2cm}

Let a metric $G$ on the torus $T^2$ is pseudo-liouville. That is
 there exists a
covering
 $\rho:S_1\times S_L\to T^2$
  such that the metric  $\rho^*G$ is given by the formula
  $(f(x)+h(y))(dx^2+dy^2)$.

Consider the standard plane $R^2$ with the standard coordinates $x,y$.
Denote by $\xi$  the action  of the group $Z\times Z$,
generated by shifts along the
vectors $(1,0)$ and $(0,L)$.
We may consider the torus $T_L$ as the factor-space of $R^2$ by $\xi$.
Since the action $\xi$ preserve coordinate lines we may use the same notation for
the coordinates on $R^2$ and $T_L$.

Consider the universal covering $U:R^2\to T_L$ that is dual to the action
$\xi$.  The mapping $U  g:R^2\to T^2$ is a universal covering or $T^2$.
The metric $(Ug)^*(G)$ on the plane $R^2$ is
given by the formula
  $(f(x)+h(y))(dx^2+dy^2)$.
Consider the action of the fundamental group $\pi_1(T^2)$ on the plane $R^2$.
Since the action is free and since it preserves angles and orientation, we see
 that the group $\pi_1(T^2)$ acts by shifts.

     Evidently, the group $\pi_1(T^2)$ is isomorphic to   $Z\times Z$.

Let the generators of the group
$Z^2$    act by shifts along  vectors
$\bar v=(v_1,v_2)$ and
$\bar u=(u_1,u_2)$.   Using lemma  1,
$f(x+u_1)=f(x+v_1)=f(x)$ and
                  $h(y+u_2)=h(y+v_2)=h(y)$. Let us prove that the fractions
                  $\frac{v_1}{u_1}$  and
$\frac{v_2}{u_2}$  are rational numbers.

Assume the converse. Then
 for any positive number $\epsilon$ there exists the pair
  of  integer numbers
 $n_1,n_2$ such that $0<n_1v_1+n_2v_2<\epsilon$.  Therefore the derivative of
 $f$ is equal to zero in every point. Hence,
  $f\equiv const$. Contradiction.

Thus there exist the numbers $\alpha_1$, $\alpha_2$ and the $2\times2 $
matrix $A$ of integers such that $(\alpha_1,0)=uA$, $(0,\alpha_2)=vA$.
Therefore,
$f(x)=f(x+\alpha_1)$,
$h(y)=h(y+\alpha_2)$. We see that
 shifts along the vectors $(\alpha_1,0)$ and $(0,\alpha_2)$
preserve the metric
$(f(x)+h(y))(dx^2+dy^2)$.

Consider the action of the group $Z\times Z$, generated by shifts along vectors
$(\alpha_1,0)$ and $(0,\alpha_2)$. Denote by  $T_l$ the factor-space of $R^2$ by
this action. Since the action preserves the
metric $(f(x)+h(y))(dx^2+dy^2)$, we see that the metric $(f(x)+h(y))(dx^2+dy^2)$ induces the metric
on $T_l$.   It is clear that the metric on the torus $T_l$ is Liouville,
and the induced mapping $T^2\to T_l$ is covering. The theorem is proved.

\vspace{.2cm}
\vspace{.2cm}

{ \bf \S3. Quadratically integrable geodesic flows on the Klein bottle.
}

Suppose $L$ is a positive number, $f,h:R\to R$  are smooth positive functions
such that $f$ is non-constant   and periodic with the period $\frac{1}{2}$,
$h$ is periodic with period $L$ and
even. Consider the metric
$ds^2=(f(x)+h(y))(dx^2+dy^2)$
on the plane $R^2$.
Consider the vectors  $u=(1,0)$, $v=(0,L)$.

Denote by $s$ the "slipping reflection" $(x,y)\to(x+\frac{1}{2},-y)$.
Denote by  $\Gamma$ the group, generated by shifts along  $u$,
$v$, and by    $s$.
The    group $\Gamma$ acts freely and preserves the metric.

Consider the factorspace
 $R^2/_\xi$. It is homeomorphic to the Klein bottle.
 Actually, since $s$ changes the
 orientation, we see that $R^2/_\xi$ is nonorientable.
  Since the action preserves the vector field $(1,0)$, we see that
  on $R^2/_\xi$ there exists a vector   field without critical points.
  Hence the Euler characteristic of $R^2/_\xi$ is equal to zero.

  The metric
$(f(x)+h(y))(dx^2+dy^2)$ induces the metric
 (we denote the indused metric by
 $G_{(L,f,h)}$) on the Klein bottle
$R^2/_\xi$.

{\begin{te2}
{\it  The geodesic flow of a metric $G$ on the Klein bottle $K^2$ is
quadratically integrable iff for the appropriate triple $(L,f,h)$
there exists a diffeomorphism $K^2\to R^2/_\xi$ that  takes the metric $G$ to
the
metric $G_{(L,f,h)}$.
}
\end{te2}}

{\bf Remark} {\small
Two different triples can specify the  same metric.}

Consider the set of triples $(L,f,h)$.

 {Consider the following operation on the
set of triples.
      $$ {\alpha}_{\varepsilon} (L,f,h)\eqdef (L,\hat{f},h),\ \
     \mbox{where}\ \ \ \hat{f}(x)=f(x+\varepsilon)$$
     $$ \beta(L,f,h)\eqdef (L,f,\hat{h}),\ \
     \mbox{where}\ \ \ \hat{h}(y)=h(y+\frac{L}{2}) $$ $$
     \gamma(L,f,h)\eqdef (L,\hat f,h),\ \ \mbox{where}\ \ \hat{f}(x)=f(-x) $$ $$
     \delta(L,f,h)\eqdef (L,\hat{f},\hat{h}), \ \ \mbox{where}\ \ \
     \hat{f}(x)=f(x)+\mbox{const},\ \ \hat{h}(y)=h(y)-\mbox{const}.$$

                    }

{\begin{te2}
{There exists a diffeomorphism
$\chi:R^2/_\xi\to R^2/_{\hat\xi}$ that takes the metric
$G_{(L,f,h)}$ to the metric
$G_{(\hat L,\hat f,\hat h)}$ iff
 the triple
  $(L,f,h)$ could be transformed to the triple $(\hat{L},\hat{f},\hat{h})$
 by operations $\alpha$, $\beta$, $\gamma$ and $\delta$.}
 \end{te2}}

\def\пр{\mbox{$RP^2$}}

\vspace{0.2cm}

\vspace{0.2cm}

\vspace{0.2cm}

{\bf
\S4. The proof of the classification theorems for quadratically integrable geodesic flows
on the Klein bottle.}

To proof Theorem 3 and Theorem 4, we need several lemmas.
                                            Let
                                            $R^2$ be the
                                             standard plane
with the standard coordinates $x,y$, let
                                            $\lambda:R^2\to R$
be a smooth positive function.

{ \begin{le2}  Suppose  the geodesic flow of the metric
$\lambda(x,y)(dx^2+dy^2)$
on the plane $R^2$ admits  a quadratical in momenta integral.
Let the function  $\lambda$ and the coefficients of the
 integral are bounded functions.
Then    in the appropriate (global) coordinates
$\hat x$, $\hat y$ the metric
$\lambda(x,y)(dx^2+dy^2)$ is given by the formula
 $(f(\hat x)+h(\hat y))(d\hat x^2+d\hat y^2)$,
 and the integral is given by the formula
   $\frac{(f(\hat x)-C_0){\hat {p_y}}^2-(h(\hat y)+C_0){\hat p_x}^2}{f(\hat x)+h(\hat y)}$,
   where
   $C_0$ is the appropriate constant.
       \end{le2}

\vspace{.2cm}

Proof. The Hamiltonian of the geodesical flow is the function
        $ H(x,y,p_x,{p_y})=\frac{p_x^2+{p_y}^2}{\lambda(x,y)}$.
Let the integral is given by the formula
   $F(x,y,p_x,{p_y})=a(x,y)p_x^2+b(x,y)p_x{p_y}+c(x,y){p_y}^2$.
   Consider the complex-valued function
      $R_{x,y}(z)=a(x,y)-c(x,y)+ib(x,y)$ of the complex variable
   $z=x+iy$.
   Suppose the  coordinates
$(\hat x,\hat y)$  as functions of  coordinates
    $(x,y)$ are given by the formula
   $\hat z=Z_0z$, where  $Z_0$ is the appropriate complex constant,
      $ z\eqdef x+iy$,  and
   $\hat z\eqdef \hat x+i\hat y$. Let in  the
   coordinates
    $(\hat x,\hat y)$ the   integral
   $F$ is given by the formula
   $F(\hat x,\hat y,\hat p_x,\hat {p_y})=\hat a(\hat x,\hat y){\hat p_x}^2+\hat b(\hat x,\hat y)\hat p_x\hat {p_y}+\hat c(\hat x,\hat y)\hat{{p_y}}^2$,
   where $\hat p_x$ and   $\hat {p_y}$ are canonically correspondent  to
   $\hat x$, $\hat y$ momenta. Consider the function $R_{\hat x,\hat y}$,
   $R_{\hat x,\hat y}(z)\eqdef
   \hat a(\hat x,\hat y)-\hat c(\hat x,\hat y)+i\hat b(\hat x,\hat y)$.
    In          \cite{Kol1}, V. Kolokoltzov proved that functions
         $R_{x,y}$ and  $R_{\hat x,\hat y}$ are holomorfic functions of the complex
         variable $z$, and satisfy the formula
   $R_{x,y}=Z_0^2R_{\hat x,\hat y}$.

Since function
 ${R}_{x,y}$ is bounded, it follows  that it is constant.
We shall    prove that $R_{x,y}\neq 0$.

                   Assume the converse.
We have $\{ H,F\} =0$.   Therefore we have
$$\{ H,F\}=H_{p_x}F_x+H_{p_y}F_y-F_{p_x}H_x-F_{p_y}H_y=$$
$$=2\lambda^{-2}((a_x\lambda
+a{\lambda}_x)p_x^3+(a_y\lambda+a{\lambda}_y)p_x^2{p_y}+
(a_x\lambda +a{\lambda}_x)p_x{p_y}^2+(a_y\lambda+a{\lambda}_y){p_y}^3)=0.$$

Since the   left part is the uniform polynom in momenta, we obtain the system
       $ \cases {{(a\lambda)}_x=0 \cr     {(a\lambda)}_y=0 \cr} $.

Thus for the appropriate constant $D$ \
we have        $a=\frac{D}{\lambda(x,y)} $.

       If we replace   $a$ by $\frac{D}{\lambda(x,y)}$ in the formula for the 
 integral $F$,
        we obtain  $F=DH$. The last formula  contradicts the functional independence
        of $H$ and $F$.

For the appropriate constant $Z_0$ the function  $R_{\hat x,\hat y}\equiv 1$.
Therefore, $\hat b(\hat x,\hat y)\equiv 0$ and
$\hat a(\hat x,\hat y)=\hat c(\hat x,\hat y)+1$.

The condition
$\{ H,F\}=0$ implies the following system:
        $  \cases { {({\hat{a}(\hat x,\hat y)}\lambda)}_{\hat x}=0  \cr
{({\hat{a}(\hat x,\hat y)}\lambda)}_{\hat y}={\lambda}_{\hat y} \cr }$.

Hence,
${(a\lambda)}_{xy}-{\lambda}_{xy}-{(a\lambda)}_{yx}=0$. Therefore,
${\lambda}=f(x)+h(y)$.      Substituting  $f(x)+h(y)$
for  $\lambda$ in the system, we obtain \
$\hat a(\hat x,\hat y)=\frac{f(\hat x)-C_0}{f(\hat x)+h(\hat y)}$,
$\hat b\equiv 0$, and
$\hat c(\hat x,\hat y)=-\frac{h(\hat y)-C_0}{f(\hat x)+h(\hat y)}$.

Thus,
   $F(\hat x,\hat y,\hat p_x,\hat {p_y})=\frac{(f(\hat x)-C_0){\hat {p_y}}^2-(h(\hat y)+C_0){\hat p_x}^2}{f(\hat x)+h(\hat y)}$.
   Proved.

\vspace{.2cm}

{\begin{le2} Suppose the  function $\lambda$ is non-constant, bounded, and
${\lambda}_{xy}={\lambda}_{\hat x\hat y}=0$. If coordinates
$(\hat{x},\hat{y})$ as functions of coordinates  $(x,y)$  are given by either
formula
$\hat z=Z_0z+Z_1$ or  $\hat{z}=Z_0\bar{z} +Z_1$, then the number $Z_0$ is pure real or pure imaginary
number.
            \end{le2}}

\vspace{.2cm}

Proof. Suppose  coordinates
$(\hat{x},\hat{y})$   are given by the formula
$\hat z=Z_0z$. Suppose   $Z_0$ is equal to $re^{i\phi}$.
Then,
$\hat x=r(x\cos{\phi}+y\sin{\phi})$, $\hat y=r(-x\sin{\phi}+y\cos{\phi})$.
Therefore,
${\lambda}_{\hat x\hat y}=
r^2\cos{\phi}\sin{\phi}({\lambda}_{xx}-{\lambda}_{yy})=0$.
We shall prove that
${\lambda}_{xx}-{\lambda}_{yy}\ne 0 $. Assume converse.
Since $\lambda_{xy}=0$, we see that
${\lambda}_{xx} $    depends only on
 $x$,
${\lambda}_{yy} $ depends only on
 $y$. Therefore,
${\lambda}_{xx}={\lambda}_{yy}=\mbox{const}$.
Since
${\lambda} $  is bounded, we see that
${\lambda}=\mbox{const}$. Contradiction.

We have
$r^2\cos{\phi}\sin{\phi}({\lambda}_{xx}-{\lambda}_{yy})=0$. Hence,
$r^2\cos{\phi}\sin{\phi}=0$. Proved.

           \vspace{3mm}

 {\bf   Corollary }  {\it \small Suppose   $\lambda:R^2\to R$  is bounded non-constant function,
 diffeomorphism  $\chi:R^2\to R^2$   preserves the metric
 $\lambda(x,y)(dx^2+dy^2)$, and  the geodesic flow of the metric admits a
 quadratic in momenta integral $F$. Then for any integral $F_1$ for the appropriate
 constants $C_0$, $C_1\ne 0$  \ $F_1=C_0H+C_1F$.        }

           \vspace{3mm}

 Let us prove Theorem   3. Suppose the geodesic flow of the metric $G_K$
 on the Klein bottle  $K^2$
 admits a quadratic in momenta integral $F_K$.
Consider the simple-connected covering  $U:R^2\to K^2$.  Denote by $G$ the
metric $U^*(G_K)$ on the plane $R^2$; by $F$ we denote
 the function $U^*(F_K)$.
Evidently,  $F$ is a quadratic in momenta  integral of the geodesical flow of the metric $G$.

It is known that in the appropriate coordinates $x,y$ the
metric $G$ is given by the
formula   $\lambda(x,y)(dx^2+dy^2)$, where
$\lambda:R^2\to R$ is  positive function.

 Since the Klein bottle is a compactum  set, we see that the function
 $\lambda$ and the coefficients of the integral
          $F$ are bounded.

  It is clear that representation of the fundamental group  $\pi_1(K^2)$ is the following:
$<a,b||abab^{-1}=1>$.
We shall describe the action of the  group  $\pi_1(K^2)$  on  $R^2$.

It is easy to proof that the action of
an arbitrary element $c\in \pi_1(K^2)$ is given  either by formula
 $Az+B$ or  $A\bar z+B$, where  $A$ and
$B$ are the appropriate complex constants, and $z\eqdef x+iy$  is the
complex coordinate.
Indeed,   any  $a\in \pi_1(K^2)$  acts by conformal auto-diffeomorphism
$R^2\to R^2$.

Let us show that the complex constant $A$ is equal to 1.
First  since the action
     preserves the volume, we have  $|A|=1$.
     Secondly since the action is free, we see that $A=1$.

Thus, the oriented elements of $\pi_1(K^2)$ act by shifts,
non-oriented elements act by
slipped reflections.

Consider the subgroup $<a,b^2> $ of $\pi_1(K^2)$,
generated by the elements $a$ and $b^2$.
The elements $a$ and $b^2$ are oriented. Therefore they act by shifts. Denote by
$\bar a$, $\bar b$ the shifts,   correspondent to $a$ and $b^2$.
Let us prove that the vector $\bar a$ is not  parallel to the vector $\bar b$.
Assume the converse.  Then the
factorspace of   $R^2$ by the action of the subgroup
$<a,b^2>$ is not homeomorphic to the torus. Contradiction.

Let us show that     the element $b$ acts by the slipping reflection
with the directing vector $\frac{\bar b}{2}$.
Since element  $b$ is non-oriented, we see that  it acts by a slipping reflection.
Since $b^2$ acts                     by shift along the vector
$\bar b$, we see that the directing vector is  $\frac{\bar b}{2}$.

We shall prove that the vector
$\bar a$  is orthogonal to the vector  $\bar b$. We have  $abab^{-1}=1$.
Hence,
$bab^{-1}=a^{-1}$. Let  the element $b$ acts by composition of shift along
$\frac{\bar b}{2}$ and a reflection $s_1$. Since $s_1b=bs_1$ and since
$ab=ba$, we have
$s_1as_1^{-1}=a^{-1}$. Therefore the axis of the reflection
 $s_1$ is orthogonal to
the vector  $\bar a$.   Proved.

Using lemma 2, there exists a coordinate system $x,y$, in which    the metric
 is given
by the formula  $(f(x)+h(y))(dx^2+dy^2)$.

Let we prove that the vector  $\bar b$ is parallel to
a coordinate axis of this coordinate system. Using lemma  3,
we have two cases. 1. The vector  $\bar b$ is parallel to an axis.
2. The angles between the vector $\bar b$ and the axes are equal to 
$\frac{\pi}{2}$.

We prove that the second case is impossible. Using lemma 3, we have
$f(x+{k_b})=h(x)$. Since $F$ is giving by the formula
  $\frac{(f(x)-C_0){{p_y}}^2-(h(y)+C_0){p_x}^2}{f(x)+h(y)}$,
  we see that
  $b$ does not preserves the  integral  $F$. Proved.

Thus the Klein bottle $K^2$ can be considered
as the factorspace  of $R^2$ by
the action of the group, generated by a shift along
 the first coordinate axis and
a slipped reflection along the second coordinate axis.

To complete the proof we shall prove that
$f\equiv \mbox{const}$
iff
the geodesical flow of metric $G_{(L,f,h)}$ is linear integrable.

It is clear that if
$f\equiv \mbox{const}$  then $F\eqdef p_x$ is an integral.

Let us prove that for $f\ne \mbox{const}$
the geodesic flow of the metric
$G_{(L,f,h)}$ is not linear integrable. Suppose
 a linear integral is given by the
formula $a(x,y)p_x+b(x,y)p_y$. Using lemma 2,  we have
  $$                        (a(x,y)p_x+b(x,y)p_y)^2
  =\frac{(f(x)-C_0){{p_y}}^2-(h(y)+C_0){p_x}^2}{f(x)+h(y)}.$$

Therefore, $2(f(x)-C_0)(h(y)+C_0)=0$. By assumption,
$f\ne \mbox{const}$.
Hence, $h\equiv C_0$. Then, $b\equiv \mbox{const}$, $a\equiv 0$.
But  since the slipping reflection $(x,y)\to (x+\frac{1}{2},-y)$
changes the sign of  the function $p_y$, we see that
$\mbox{const}\equiv 0$. The
theorem  is
proved.

\end{document}